\documentclass[twoside,symmetric]{tufte-handout}

\usepackage{graphicx}
\usepackage{dcolumn}
\usepackage{bm}
\usepackage[separate-uncertainty=true]{siunitx}
\usepackage{mhchem}
\usepackage{amsmath}
\usepackage{hyperref}

\begin{document}

\title{Neutron reflectometry analysis: using model-dependent methods}

\author{%
    Andrew R. McCluskey\thanks{Diamond Light Source \& University of Bath. email: \href{mailto:andrew.mccluskey@diamond.ac.uk}{andrew.mccluskey@diamond.ac.uk}}}
  
\date{\today}

\maketitle

\begin{abstract}
\noindent
Neutron reflectometry analysis is an inherently ill-posed, which is to say that there are many possible solutions which agree equally well with the measured data. 
This leads to the application of model-dependent analysis, where information that we \textbf{know} about the system is integrated into our analysis. 
This tutorial briefly covers the mathematics underlying the use of model-dependent analysis in neutron reflectometry. 
\end{abstract}

\section{Introduction}

The aim of model-dependent neutron reflectometry analysis is to create a \textbf{model} that accurately reproduces the collected data.\cite{lovell_analysis_1999} 
The collected data (once normalised) is described in terms of \textbf{reflected intensity}, $R(q)$, which depends on the wavevector, $q$, and is measured as, 
\begin{equation}
    R(q) = \frac{\text{specular reflected neutron intensity at $q$}}{\text{incident neutron intensity}},
    \label{equ:refl}
\end{equation}
where the denominator in Equation~\ref{equ:refl} is the total neutron flux on the sample.
It is this $R(q)$ that we want to calculate from some model and compare with our data. 

\begin{figure}
    \forceversofloat
    \includegraphics[width=\textwidth]{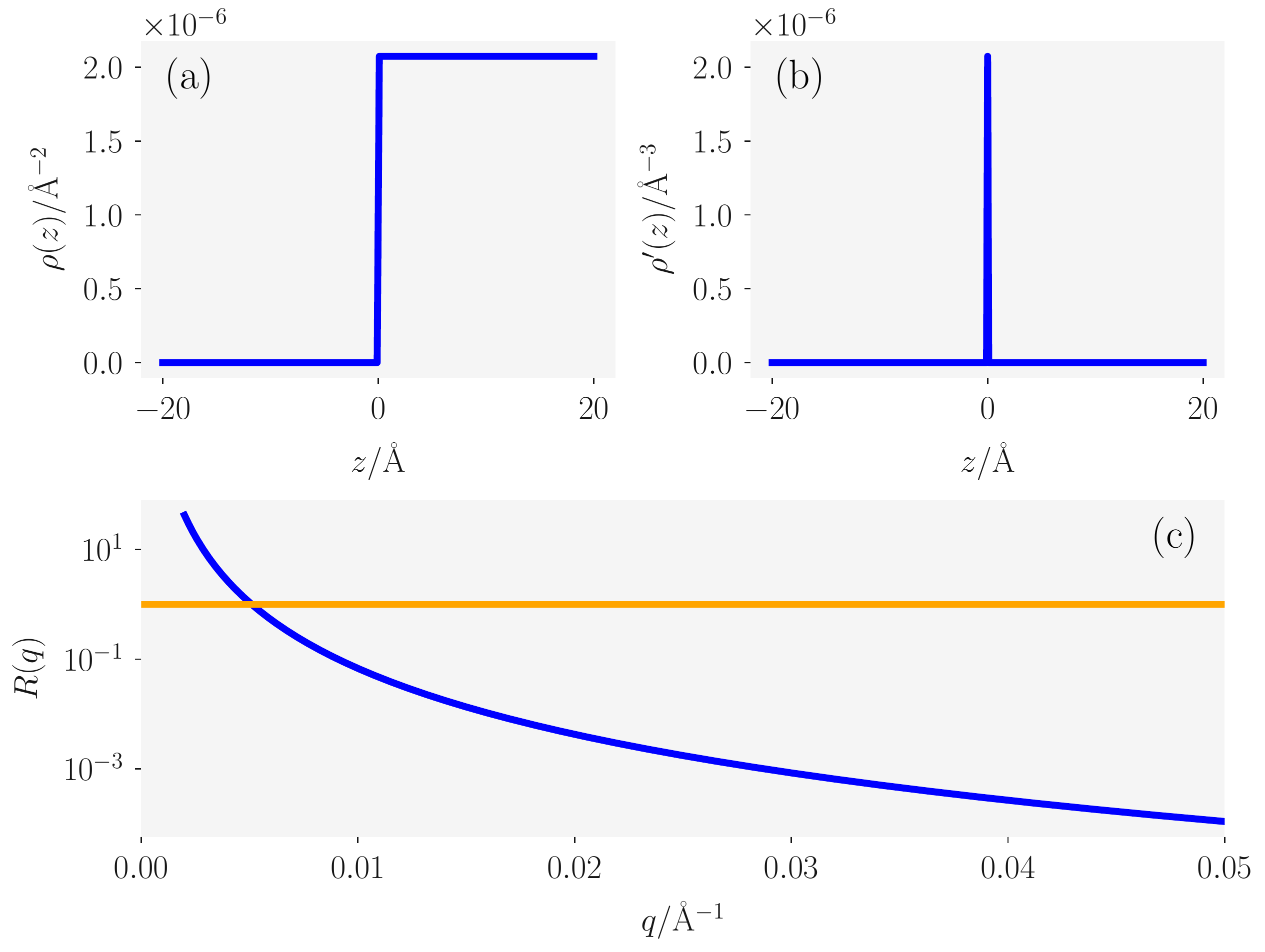}
    \caption{A graphical representation of the kinematic approach; (a) the Heaviside function describing the scattering length density profile for a bare silicon substrate, (b) the $\delta$-function arising from the first derivative of the function in (a), and (c) the resulting reflectometry profile, where the orange line at $R(q)$ = 1 identifies the breakdown of between experiment and theory.}
    \label{fig:kine}
\end{figure}

By applying the \textbf{Born approximation},\cite{born_quantenmechanik_1926} that there is only a single scattering event for each neutron, the reflected intensity can be approximated as follows,\cite{sivia_elementary_2011}
\begin{equation}
    R(q) \approx \frac{16\pi^2}{q^4} \bigg| \int^{+\infty}_{-\infty}{\rho'(z)\exp{(-\mathrm{i} zq) \text{d}z} \bigg|^2},
    \label{equ:kine}
\end{equation}
where, $\rho'(z)$ is the first derivative of the scattering length density profile.\footnote{The scattering length density profile is our model.} 
However, this \textbf{kinematic} approach has a significant problem, which can be demonstrated by considering a the simple case of a bare silicon substrate, which can be modelled with a Heaviside function, shown in Figure~\ref{fig:kine}(a), 
\begin{equation}
    \rho(z) = 
    \begin{cases}
        0 & \text{where } z < 0,\\
        \rho_{\text{Si}} & \text{otherwise},
    \end{cases}
\end{equation}
where, $\rho_{\text{Si}} = \SI{2.074e-6}{\angstrom^{-2}}$. 
The first derivative of a Heaviside function is a scaled $\delta$-function, shown in Figure~\ref{fig:kine}(b), 
\begin{equation}
    \rho'(z) = \rho_{\text{Si}}\delta(z).
\end{equation}
Then, using Equation~\ref{equ:kine}, we can calculate the reflected intensity with respect to $q$, 
\begin{equation}
    \begin{aligned}
    R(q) & \approx \frac{16\pi^2}{q^4} \bigg| \rho_{\text{Si}}\int^{+\infty}_{-\infty}{\delta(z)\exp{(-\mathrm{i} zq) \text{d}z}} \bigg|^2 \\ 
     & \approx \frac{16\pi^2}{q^4} \bigg| \rho_{\text{Si}} \exp{(0)} \bigg| ^2 \\
     & \approx \frac{16\pi^2\rho_{\text{Si}}^2}{q^4}.
    \end{aligned}
\end{equation}
This reflectometry profile is shown in Figure~\ref{fig:kine}(c), where it is clear that the agreement with the experimental profile would be poor as $q \to 0$.\cite{majkrzak_exact_1998}
For low values of $q$, the model reflectometry from this simple system is greater than \num{1}, violating the physical condition imposed by Equation~\ref{equ:refl}\footnote{In the kinematic approach, more neutrons are reflected than are indicent in the first place!}.
This breakdown of the kinematic approach is due to the assumption of the Born approximation, that each neutron is only scattered once. 
The reflectometry geometry, with the large path length that arises the shallow incidence angle, means that multiple scattering events are likely, rendering the kinematic approach invalid.

\section{Recursive methods}

The breakdown of the kinematic approach has lead to the application of the Abel\`{e}s,\cite{abeles_sur_1948} or Parratt,\cite{parratt_surface_1954} recursive model for the reflection of light at a given number of stratified interfaces.\footnote{This method is also known as the dynamical approach.}
This is an alternative method for calculating the reflected intensity from a model and involves describing the system as a series of layers of different scattering length density. 
The neutrons can be either reflected or refracted at the interface between each of the layers.
\begin{figure}
    \includegraphics[width=\textwidth]{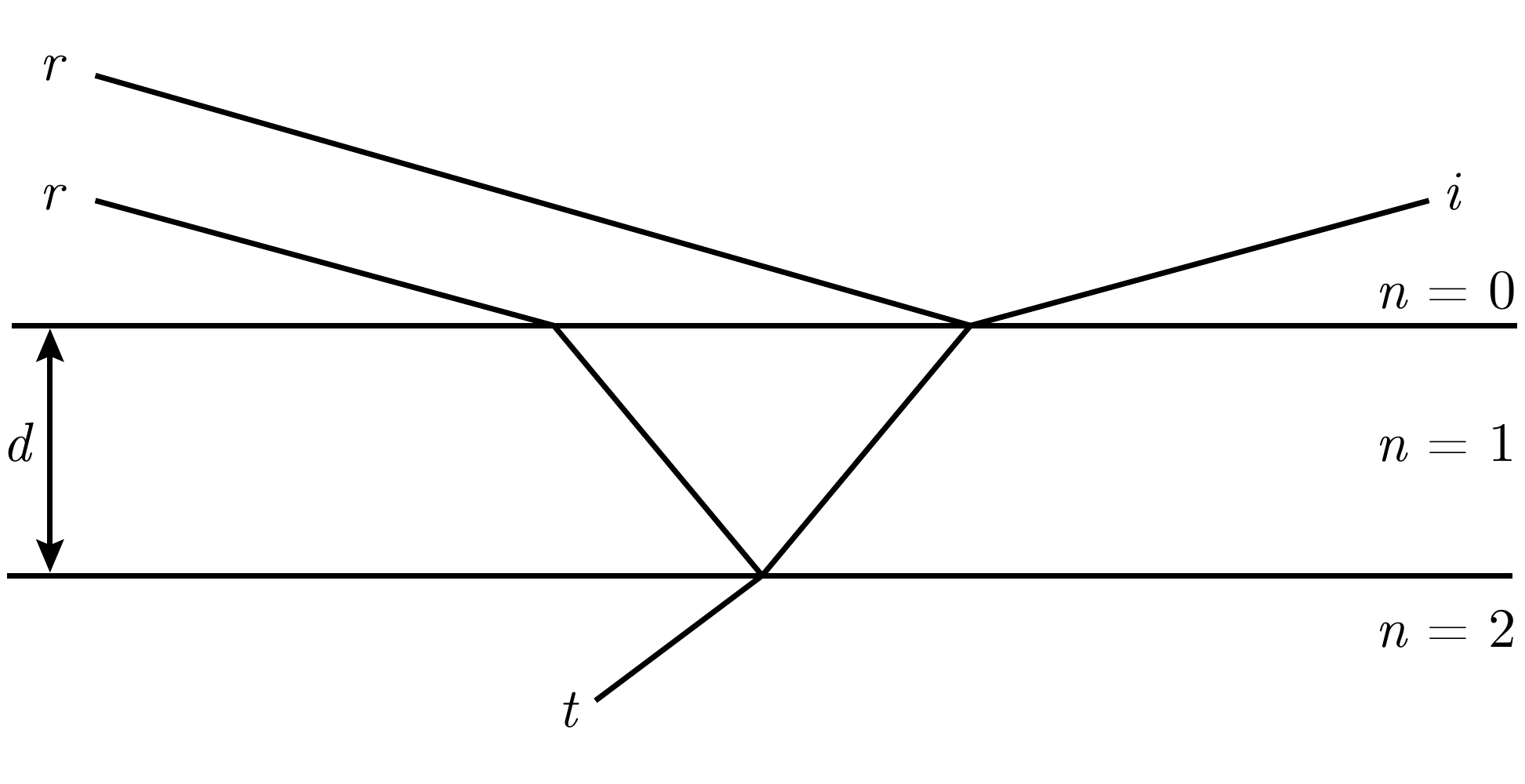}
    \caption{A schematic diagram showing the reflected ($r$) and transmitted ($t$) waves when an incident ($i$) wave enters an interface of thickness $d$.}
    \label{fig:refr}
\end{figure}
Figure~\ref{fig:refr} shows the refraction and reflections from a simple two layer system, where layer \num{0} is a vacuum above the sample. 
It is clear how the two waves, labelled $r$, could interfere constructively or destructively depending on the thickness of layer \num{1}, $d$.

The generalisation of this approach to any number of layers is possible and enables the reflected intensity to be calculated at each value of $q$ that was measured. 
The incident neutrons will be refracted by each of the layers, giving wavevectors for each layer, $k_n$, 
\begin{equation}
    k_n = \sqrt{k_0+4\pi(\rho_n - \rho_0)},
\end{equation}
where, $k_0 = 0.5q$. 
The Fresnel equation coefficient between layers $n$ and $n+1$, $r_{n, n+1}$i, can then be found along with the phase factor, $\beta_n$, for the layer $n$, 
\begin{equation}
    r_{n, n+1} = \frac{k_n - k_{n+1}}{k_n + k_{n+1}}, 
    \label{equ:fres}
\end{equation}
\begin{equation}
    \beta_n = k_nd_n.
\end{equation}
From the Fresnel equation coefficient and phase factor, the characteristic matrix for each layer, $M_n$, can be constructed, 
\begin{equation}
    M_n = 
    \begin{matrix}
        \exp{\beta_n} & r_{n,n+1}\exp{-\beta_n} \\ 
        r_{n,n+1}\exp{\beta_n} & \exp{-\beta_n}
    \end{matrix},
\end{equation}
and the resultant matrix $B$ is found from the product sum of the matrices from each layer, 
\begin{equation}
    B = \prod_{n=0}^{n_{\text{max}}}{M_n}.
\end{equation}
The final model reflected intensity at a given value of $q$ is found from the following elements of the resultant matrix, 
\begin{equation}
    R(q) = \frac{B_{1,2}}{B_{1,1}}. 
\end{equation}

\begin{figure}
    \includegraphics[width=\textwidth]{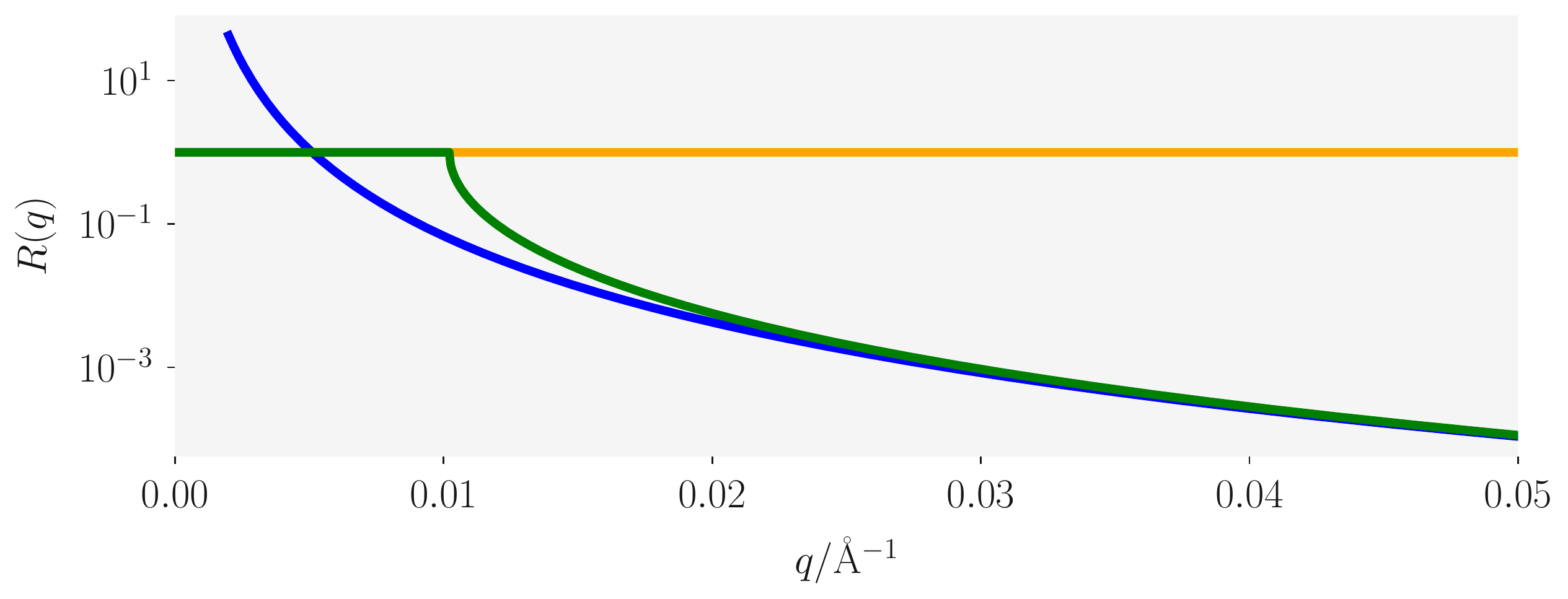}
    \caption{A comparison of the kinematic (blue) and dynamical (green) approaches to determine the reflected intensity from the material with the scattering length density profile given in Figure~\ref{fig:kine}(a).}
    \label{fig:dyna}
\end{figure}

This algorithm models the layers as \textbf{perfectly} flat layers, which will not be strictly true. 
This has resulted in the use of correction terms to be added to Equation~\ref{equ:fres} to account for this \textbf{roughness}. 
The most common of these is N\'{e}vot and Croce Gaussian broadening,\cite{nevot_caracterisation_1980} in which the Fresnel equation coefficient is evaluated as, 
\begin{equation}
    r_{n, n+1} = \frac{k_n - k_{n+1}}{k_n + k_{n+1}} \exp{(-2k_nk_{n+1}\sigma^2_{n,n+1})},
\end{equation}
where, $\sigma_{n, n+1}$ is the interfacial roughness between the layers $n$ and $n+1$. 
Applying this \textbf{dynamical approach} is shown in Figure~\ref{fig:dyna}, where there is a clear difference between the kinematic and dynamical approaches as $q\to0$, with the dynamical approach adhering to the physical constraint that causes the breakdown of the kinematic approach.  

\section{Global optimisation}

The recursive method described above gives an accurate method to obtain a model reflected intensity. 
However, this is just the first step in the analysis of a neutron reflectometry dataset. 
Now we are interested in optimising our model such that the reflected intensity from it matches our experimental data as best as possible. 
This is the problem of parameter optimisation, which is a broad area of mathematics and computer science that we will not dwell on here.\footnote{There are whole journals dedicated to the subject, see SIAM Journal on Optimization or Journal of Global Optimization.}
However, we will introduce the basics of optimisation and mention the most common \textbf{global optimisation} method applied in reflectometry.

\begin{figure}
    \includegraphics[width=\textwidth]{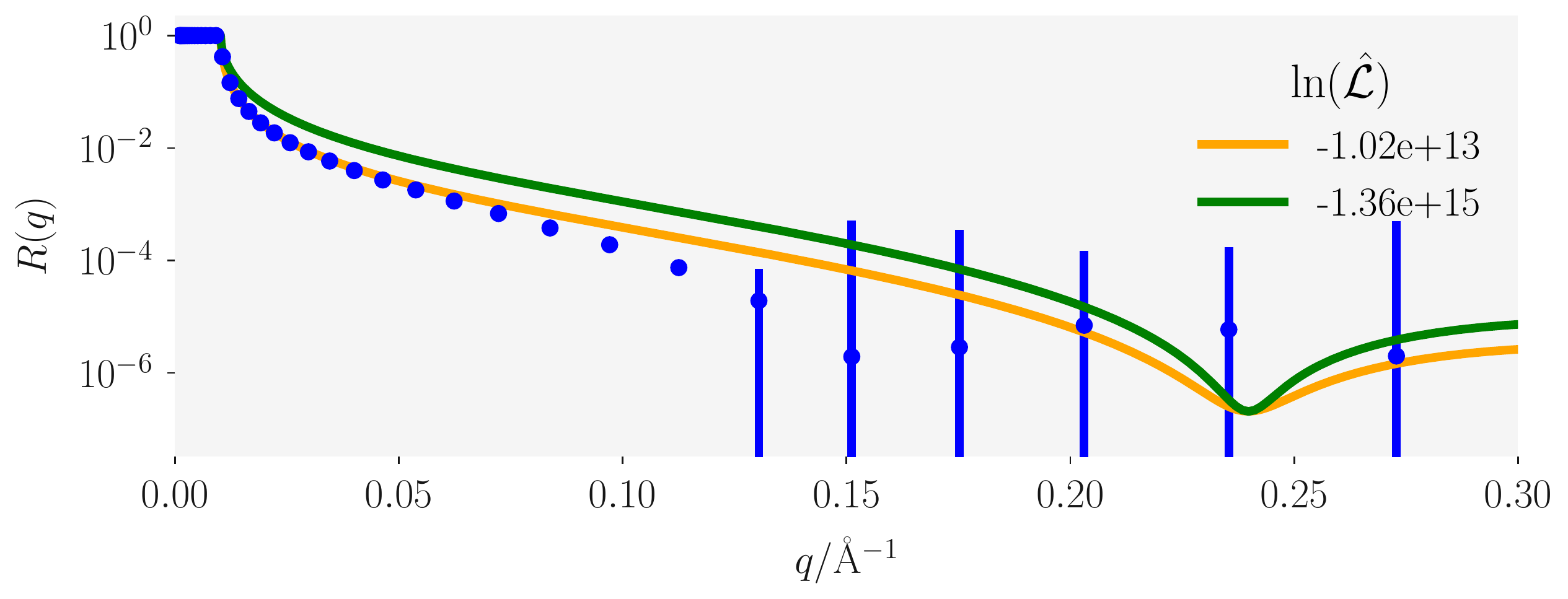}
    \caption{The maximum likelihood fit, the blue data points indicate some experimental data, while the yellow line shows the maximum likelihood model fit to this dataset and the green line show another model (which doesn't maximise the likelihood).}
    \label{fig:likelihood}
\end{figure}

When we measure a reflectometry profile, we measure the reflected intensity (and some uncertainty in that measurement) at discrete points in the wavevector, $R(q) \pm \delta R(q)$.
Using the recursive method discussed above, we can calculate the model reflected intensity at these same $q$ values, $R_m(q)$. 
We then aim to reduce the difference between the measured and modelled reflected intensity through the optimisation (maximisation) of the \textbf{likelihood} $\mathcal{L}$, 
\begin{equation}
    \mathcal{L} = \exp{\bigg\{-0.5 \sum_{q=q_{\text{min}}}^{q_{\text{max}}} \bigg[\frac{R(q) - R_m(q)}{\delta R(q)}\bigg]^2 + \ln[2\pi \delta R(q)]\bigg\}}.
\end{equation}
This parameter describes the \textbf{probability} that the model $m$ accurately describes the observed reflectometry data.
The aim in reflectometry fitting is to obtain model reflectometry at has the maximum likelihood possible for the given experimental data. 
Figure~\ref{fig:likelihood} shows the maximum likelihood model for some experimental data, in addition to another model which doesn't agree well with the data and has a lower likelihood as a result.

The global optimisation of a reflectometry model is a particularly difficult problem, this is due to the \textbf{ill-posed nature} of this data, this is where there are many reasonable solutions to a particular reflectometry profile.\footnote{This can also be thought of as \emph{every reflectometry curve looks the same}, which you may have heard before.}
However, a particular global optimisation method has shown substantial utility in the fitting of reflectometry data,\cite{varderlee_comparison_2007} \textbf{differential evolution}.\cite{wormington_characterization_1999}
This has lead to the inclusion of this method in many common reflectometry analysis packages.\cite{bjorck_fitting_2011}

Differential evolution is an iterative, genetic algorithm, designed to mimic the evolution processes observed in biology.\cite{holland_adaptation_1992}
The method consists of two vectors, the parent population $\mathbf{p}$, and the offspring population, $\mathbf{o}$. 
These vectors are of shape $(i \times j)$, where $i$ is the number of parameters in the model and $j$ is the number of candidate solutions being considered. 
The offspring population is generated as a result of some trail method.\footnote{Here we will only discuss a simple classical trail method, however, many of these exist.}

A classical trail method consists of two stages, mutation and recombination. 
The \textbf{mutation} stage involves the creation of a mutant population, $\mathbf{m}$. 
The magnitude of the mutation is dependent on the first of our \textbf{hyperparameters}, the mutation constant, $k_m$, 
\begin{equation}
    \mathbf{m}_{i,j} = b_i + k_m (\mathbf{p}_{i, R1} - \mathbf{p}_{i, R2}),
\end{equation}
where $b$ is the candidate solution with the greatest likelihood, and $\mathbf{p}_{i, R1}$ and $\mathbf{p}_{i, R2}$ are randomly chosen members of the parent population. 
The mutation constant hyperparameter controls the size of the search space, with a large $k_m$ corresponding to a wider search. 

The \textbf{recombination} step creates the offspring population vector by taking a sample from either the parent or mutant population with some frequency, which depends on our second \textbf{hyperparameter}, the recombination constant, $k_r$,
\begin{equation}
    \mathbf{o}_{i, j} = 
    \begin{cases}
        \mathbf{m}_{i, j} & \text{where } X < k_r,\\
        \mathbf{p}_{i, j} & \text{otherwise},
    \end{cases}
\end{equation}
where, $X$ is a random number selected from a uniform distribution between 0 and 1. 
The recombination constant hyperparameter controls the mutation frequency in the offspring population. 

The final stage is to compare the offspring and parent population vectors, in the \textbf{selection} stage, to create the parent population for the next iteration. 
Here, the likelihood is used to compare between subunits from the offspring or parent populations, 
\begin{equation}
    \mathbf{p}_{*, j} \leftarrow 
    \begin{cases}
        \mathbf{o}_{*, j} & \text{where } \mathcal{L}_{\mathbf{o}_{*, j}} > \mathcal{L}_{\mathbf{p}_{*, j}},\\
        \mathbf{p}_{*, j} & \text{otherwise},
    \end{cases}
\end{equation}
where, the $*, j$ subscript notation indicates all objects from the population, $j$.

\begin{figure}
    \forcerectofloat
    \includegraphics[width=\textwidth]{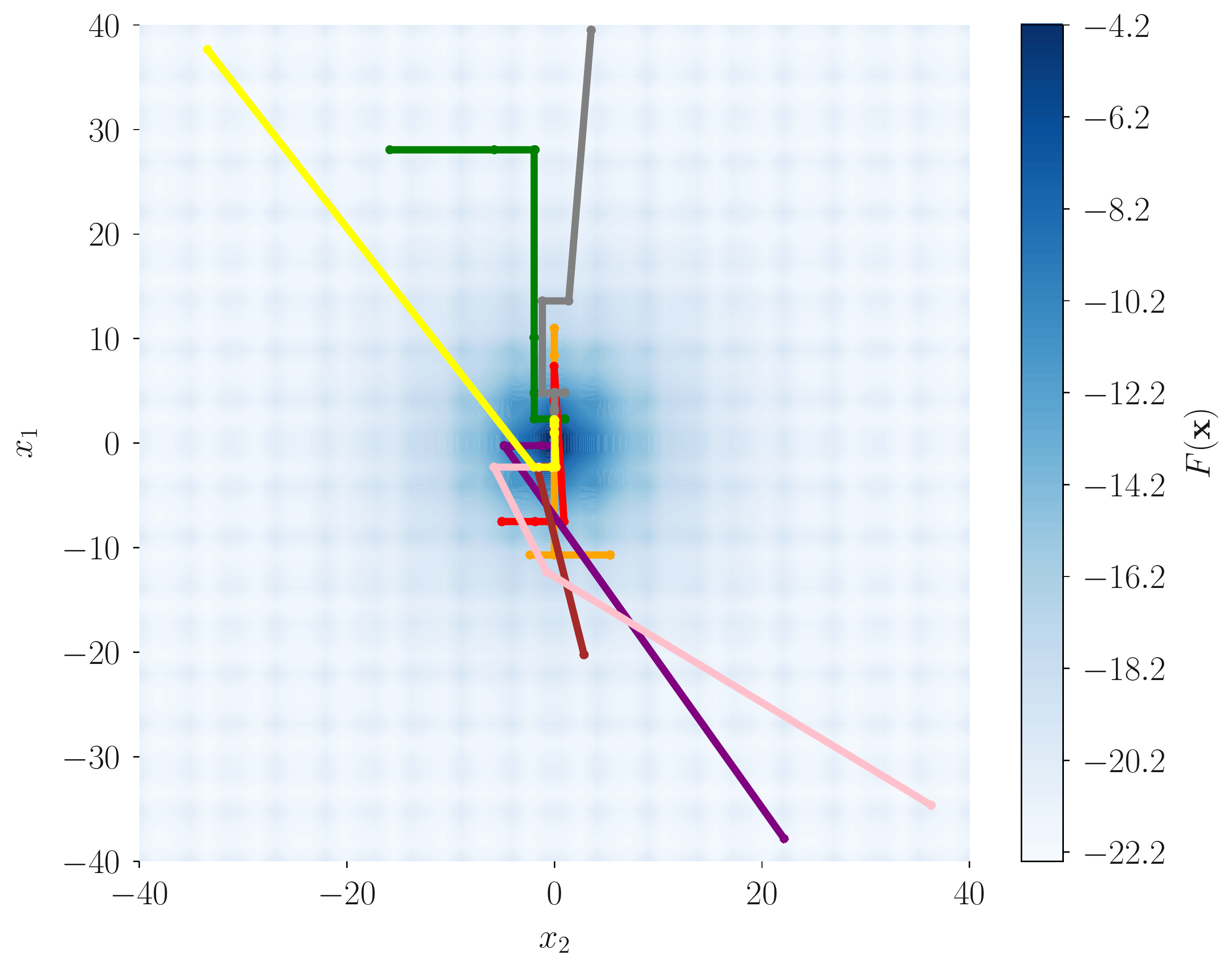}
    \caption{An example of a differential evolution algorithm applied to the negative of an Ackely function. In this implementation $k_m=0.5$ and $k_r=0.5$. Each line represents a different candidate solution. The optimisation was stopped after 100 iterations had run.}
    \label{fig:ackley}
\end{figure}

The differential evolution algorithm can be seen in action applied to the negative two-dimensional Ackley function,\cite{ackley_connectionist_1987} in Figure~\ref{fig:ackley}.\footnote{The Ackley function is a common function used in the assessment of global optimisation functions. This is due to there being a large number of local minima and only a single global minimum to this function. Here, we want to maximise the value, so the negative Ackley function is used.}
This function can maximise the value of the negative Ackley function.
While this does not offer a clear example of this algorithm's application to reflectometry analysis, the popularity of this method in the fitting of reflectometry profiles cannot be denied.\cite{bjorck_fitting_2011,nelson_refnx_2019} 

\section{Uncertainty sampling}

Reflectometry measurements offer an \textbf{average} description of the out of place structure of a material. 
This means that it is pragmatic to describe the uncertainties in the values of our model parameters in some fashion. 
This is often performed using \textbf{Markov-chain Monte Carlo} (MCMC), which is a methodology used to sample the \textbf{posterior probability distribution} for each of our parameters.\cite{sivia_data_2006}
Typically, MCMC is used on already optimised solutions to a particular problem, is in reflectometry analysis it is usually applied after the differential evolution has optimised the structure. 
In addition to being able to quantify the \textbf{inverse uncertainties},\footnote{This is the name given to the uncertainties in the model parameters.} MCMC also offers a more complete understanding of the correlations between different parameters,\cite{gilks_markov_1995} which is particularly important in the ill-posed reflectometry analysis. 

Once an optimised set of model parameters, $\theta$, are obtained which maximise the likelihood, $\mathcal{L}$, some random perturbation is applied, 
\begin{equation}
    \Theta = \theta + aR,
\end{equation}
where $R$ is some normally distributed number centered on \num{0} with a standard distribution of \num{1} and $a$ is the step size. 
A new $\mathcal{L}$ is found for $\Theta$ and the probability that this transition is will occur is found, 
\begin{equation}
    p = \exp{\bigg[\frac{\mathcal{L}()\Theta) - \mathcal{L}(\theta)}{2}\bigg]}. 
\end{equation}
This probability is then compared with a uniformly distributed random number from \num{0} to \num{1}, $n$,
\begin{equation}
    \theta \leftarrow
    \begin{cases}
        \Theta & \text{where } n < p,\\
        \theta & \text{otherwise}.
    \end{cases}
\end{equation}
This process is repeated until some desired number of samples has been obtained. 
It should be noted that it is important to allow the Markov chains to have some ``burn-in'' period, which is not included in the final samples. 
This allows the MCMC algorithm to settle into the search-space. 

\begin{figure}
    \includegraphics[width=\textwidth]{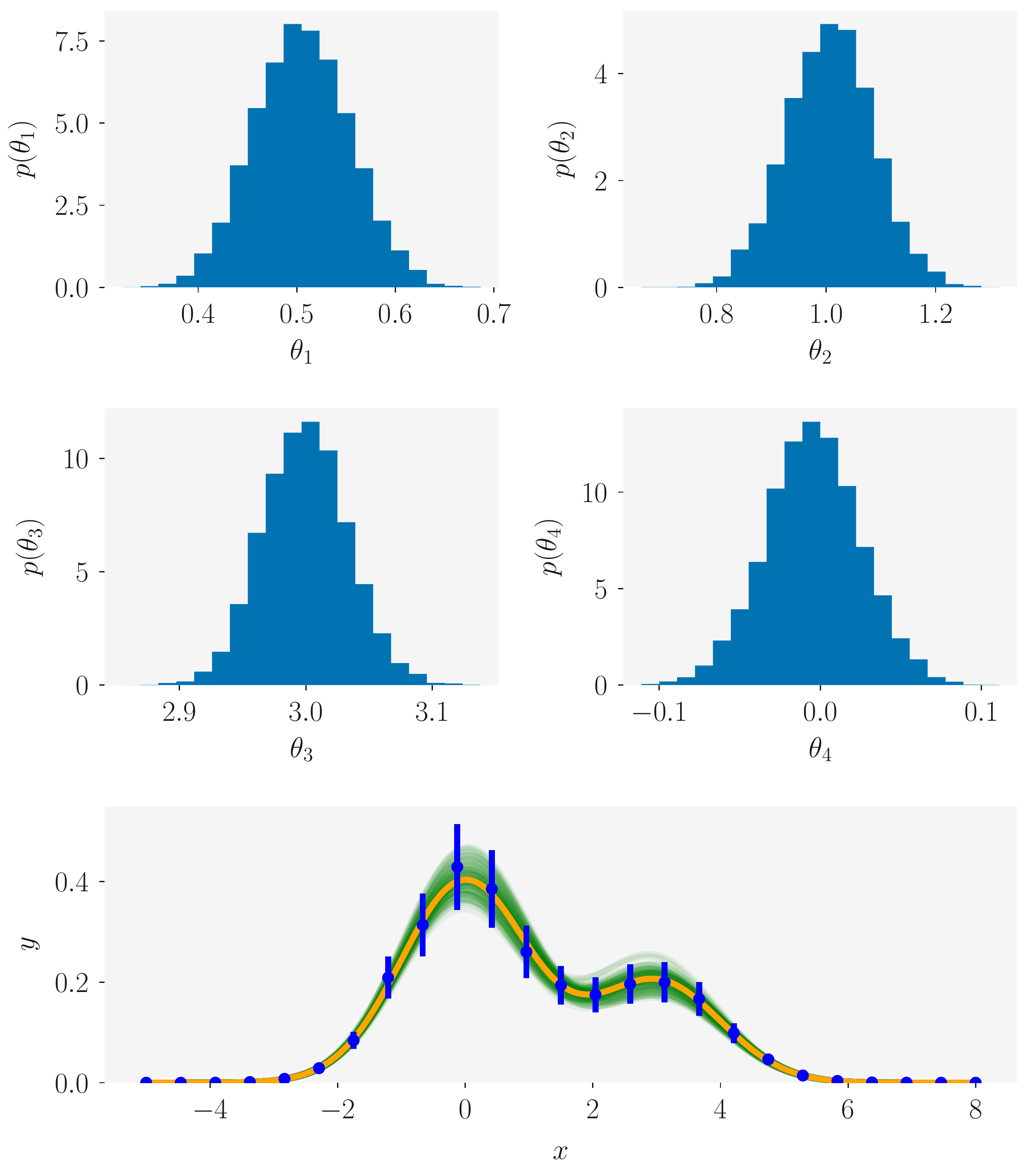}
    \caption{An example of the result of MCMC uncertainty sampling on data synthesized from two overlapping Gaussian functions. $\theta_1$ and $\theta_2$ are the integrals of the Gaussian functions while $\theta_3$ and $\theta_4$ are the positions. The data is shown with blue circles, the optimised model with an orange line, and samples from the posterior distributions with green lines.}
    \label{fig:mcmc}
\end{figure}

Figure~\ref{fig:mcmc} shows the result of an MCMC sampling for a pair of overlapping Gaussian functions, performed using the \texttt{emcee} Python package.\cite{foremanmackey_emcee_2012}
The posterior probability distributions that are available to each of the four parameters are shown along with the data, the optimised fit and a subset of models within the posterior distributions. 
All of the samples from the posterior distributions fit within the uncertainty error bars on the data.
This shows the ability for MCMC to sample the range of the distribution that is \textbf{allowed} by the experimental uncertainty. 

\section{Conclusions}

The aim of this document was to give an introduction to some of the mathematical concepts that underpin modern reflectometry model-dependent analysis. 
We have looked at how reflectometry is calculated from a layer model description of the scattering length density profile, and why the kinematic approach fails. 
Then we have discussed the importance of the differential evolution algorithm in reflectometry analysis and detailed the operation of this algorithm. 
Finally, Markov-chain Monte Carlo was introduced in the context of uncertainty quantification for model-dependent analysis, with particular importance for reflectivity discussed. 
While not exhaustive, I hope that this document will give you the confidence in understanding to look in more detail into how the analysis of reflectometry measurements are performed.

\bibliography{handout}

\begin{thebibliography}{16}
\providecommand{\natexlab}[1]{#1}
\providecommand{\url}[1]{\texttt{#1}}
\expandafter\ifx\csname urlstyle\endcsname\relax
  \providecommand{\doi}[1]{doi: #1}\else
  \providecommand{\doi}{doi: \begingroup \urlstyle{rm}\Url}\fi

\bibitem[Abelès(1948)]{abeles_sur_1948}
F.~Abelès.
\newblock \emph{Ann. Phys.}, 12\penalty0 (3):\penalty0 504--520, 1948.
\newblock \doi{10.1051/anphys/194812030504}.

\bibitem[Ackley()]{ackley_connectionist_1987}
D.~H. Ackley.
\newblock \emph{A Connectionist Machine for Genetic Hillclimbing}.
\newblock {University of Michigan}.

\bibitem[Bj\"{o}rck(2011)]{bjorck_fitting_2011}
M.~Bj\"{o}rck.
\newblock \emph{J. Appl. Crystallogr.}, 44\penalty0 (6):\penalty0 1198--1204,
  2011.
\newblock \doi{10.1107/S0021889811041446}.

\bibitem[Born(1926)]{born_quantenmechanik_1926}
M.~Born.
\newblock \emph{Z. Physik}, 38:\penalty0 803–827, 1926.
\newblock \doi{10.1007/BF01397184}.

\bibitem[Foreman-Mackey et~al.(2019)Foreman-Mackey, Farr, Sinha, Archibald,
  Hogg, Sanders, Zuntz, Williams, Nelson, de~Val-Borro, Erhardt, Pashchenko,
  and Pla]{foremanmackey_emcee_2012}
D.~Foreman-Mackey, W.~Farr, M.~Sinha, A.~Archibald, D.~Hogg, J.~Sanders,
  J.~Zuntz, P.~Williams, A.~Nelson, M.~de~Val-Borro, T.~Erhardt, I.~Pashchenko,
  and O.~Pla.
\newblock \emph{J. Open Source Softw.}, 4\penalty0 (43):\penalty0 1864, 2019.
\newblock \doi{10.21105/joss.01864}.

\bibitem[Gilks et~al.(1995)Gilks, Richardson, and
  Spiegelhalter]{gilks_markov_1995}
W.~Gilks, S.~Richardson, and D.~Spiegelhalter.
\newblock \emph{Markov {{Chain Monte Carlo}} in {{Practice}}}.
\newblock Chapman \& {{Hall}}/{{CRC Interdisciplinary Statistics}}. {CRC
  Press}, 1995.
\newblock ISBN 978-0-412-05551-5.

\bibitem[Holland(1992)]{holland_adaptation_1992}
J.~H. Holland.
\newblock \emph{Adaptation in {{Natural}} and {{Artificial Systems}}}.
\newblock {MIT Press}, 2 edition, 1992.
\newblock ISBN 978-0-262-58111-0.

\bibitem[Lovell and Richardson(1999)]{lovell_analysis_1999}
M.~R. Lovell and R.~M. Richardson.
\newblock \emph{Curr. Opin. Colloid Interface Sci.}, 4\penalty0 (3):\penalty0
  197--204, 1999.
\newblock \doi{10.1016/S1359-0294(99)00039-4}.

\bibitem[Majkrzak and Berk(1998)]{majkrzak_exact_1998}
C.~F. Majkrzak and N.~F. Berk.
\newblock \emph{Phys. Rev. B}, 58\penalty0 (23):\penalty0 15416, 1998.
\newblock \doi{10.1103/PhysRevB.58.15416}.

\bibitem[Nelson and Prescott(2019)]{nelson_refnx_2019}
A.~R.~J. Nelson and S.~W. Prescott.
\newblock \emph{J. Appl. Crystallogr.}, 52\penalty0 (1):\penalty0 193--200,
  2019.
\newblock \doi{10.1107/S1600576718017296}.

\bibitem[N\'{e}vot and Croce(1980)]{nevot_caracterisation_1980}
L.~N\'{e}vot and P.~Croce.
\newblock \emph{Rev. Phys. Appl. (Paris)}, 15\penalty0 (3):\penalty0 761--779,
  1980.
\newblock \doi{10.1051/rphysap:01980001503076100}.

\bibitem[Parratt(1954)]{parratt_surface_1954}
L.~G. Parratt.
\newblock \emph{Phys. Rev.}, 95\penalty0 (2):\penalty0 359--369, 1954.
\newblock \doi{10.1103/PhysRev.95.359}.

\bibitem[Sivia()]{sivia_elementary_2011}
D.~S. Sivia.
\newblock \emph{Elementary {{Scattering Theory}}: {{For X}}-Ray and {{Neutron
  Users}}}.
\newblock {Oxford University Press}.
\newblock ISBN 978-0-19-922868-3.

\bibitem[Sivia and Skilling(2006)]{sivia_data_2006}
D.~S. Sivia and J.~Skilling.
\newblock \emph{Data {{Analysis}}: {{A Bayesian Tutorial}}}.
\newblock {Oxford University Press}, 2 edition, 2006.
\newblock ISBN 978-0-19-856832-2.

\bibitem[van~der Lee et~al.(2007)van~der Lee, Salah, and
  Harzallah]{varderlee_comparison_2007}
A.~van~der Lee, F.~Salah, and B.~Harzallah.
\newblock \emph{J. Appl. Crystallogr.}, 40\penalty0 (5):\penalty0 820--833,
  2007.
\newblock \doi{10.1107/S0021889807032207}.

\bibitem[Wormington et~al.(1999)Wormington, Panaccione, Matney, and
  Bowen]{wormington_characterization_1999}
M.~Wormington, C.~Panaccione, K.~M. Matney, and D.~K. Bowen.
\newblock \emph{Philos. Trans. R. Soc. London Ser. A}, 357\penalty0
  (1761):\penalty0 2827--2848, 1999.
\newblock \doi{10.1098/rsta.1999.0469}.

\end{thebibliography}
\bibliographystyle{plainnat}

\end{document}